# Strain Heterogeneity in Sheared Colloids Revealed by Neutron Scattering

Kevin Chen,[a] Bin Wu,[b,*] Guan-Rong Huang,[c,d] Gaibo Zhang,[b] and Yangyang Wang[a,†]

[a] Centre for Nanophase Materials Sciences, Oak Ridge National Laboratory, Oak Ridge, Tennessee 37831, USA.

[b] Neutron Scattering Division, Oak Ridge National Laboratory, Oak Ridge, Tennessee 37831, USA.

[c] Physics Division, National Centre for Theoretical Sciences, Hsinchu 30013, Taiwan, Republic of China.

[d] Shull Wollan Centre, the University of Tennessee and Oak Ridge National Laboratory, Oak Ridge, Tennessee 37831, USA

*wub1@ornl.gov

†wangy@ornl.gov

**Recent computational and theoretical studies have shown that the deformation of colloidal suspensions under a steady shear is highly heterogeneous at the particle level and demonstrate a critical influence on the macroscopic deformation behavior. Despite its relevance to a wide variety of industrial applications of colloidal suspensions, scattering studies focusing on addressing the heterogeneity of the non-equilibrium colloidal structure are scarce thus far. Here, we report the first experimental result using small-angle neutron scattering. From the evolution of strain heterogeneity, we conclude that the shear-induced deformation transforms**

**from nearly affine behavior at low shear rates, to plastic rearrangements when the shear rate is high.**

Understanding the rheological properties of colloidal materials is a prerequisite for fully facilitating their uses in many industrial applications.[1-3] In this context, it is of great scientific as well as technological interest to establish the deformation mechanism of colloids on the particle level. Theoretical and computational investigations have been carried out extensively in the past to explore a microscopic description of the macroscopic deformation behavior.[4-12] For example, computer simulations have demonstrated a positive correlation between rheological behavior and transient heterogeneity – the spatial inhomogeneity in relaxation dynamics.[13-17] In the words, steady shear has been found to fluidize the structure by promoting local plastic flows, which are characterized by configurational fluctuations. The nonlinear shear thinning phenomenon has also been shown to be accompanied by diminishing transient heterogeneities.[14-15] Experimental investigations with imaging techniques have been performed to explore the influence of localized plastic rearrangement in the deformation behavior.[16-26] Subsequently, the spatial correlation of localized plastic events was identified and its connection to the nonlinear rheological behavior was also confirmed.[21-25]

There also exists a considerable amount of scattering studies, mainly based on small-angle scattering techniques, for investigating the non-equilibrium structure of colloids subject to steady shear.[27-31] While results from these extensive studies have significantly broadened our knowledge about the non-equilibrium microstructure of colloids under various strain conditions, the characterization of flow heterogeneity from the scattering characteristics remains unaddressed so far.

Motivated by this challenge, we investigated the non-equilibrium structure of a model colloidal system subject to steady shear using small-angle neutron scattering (SANS). Based on this framework,[32] we analyzed the anisotropic scattering spectra as a function of shear rate and demonstrated that the strain heterogeneity at the particle level can be qualitatively evaluated from the deviation of particle deformation from the ideal affine condition.

Commercially obtained silica particles (Seahostar KE-P10 Nippon Shokubai Co. Ltd.) were suspended in a mixture of ethylene glycol and glycerol. Due to the deprotonation reaction of hydroxyl groups on its surface, the silica nanoparticle becomes negatively charged. To bypass the shear-induced crystallization, a binary mixture of silica particles with diameters of 126 nm and 80 nm was used. The number density ratio of the 126 nm particles to the 80 nm particles was 4 to 1. The total volume fraction of the silica particles was set to be 0.4 and the mass ratio of the ethylene glycol to glycerol mixture was 2.27 to 1. Fig. 1 presents the results of the rheological measurements on our system, which were carried out at CNMS, ORNL using an HR2 rheometer from TA Instruments. Small-amplitude frequency sweep experiment [Fig. 1(a)] shows that the storage modulus $G'$ is much larger than the loss modulus $G''$ at all frequencies investigated. In our colloidal system consisting of rigid silica particles, the solid-like response cannot originate to any significant extent from the deformation of silica particles but is overwhelmingly due to changes in the statistical distribution of the particle positions, and thus directly linked to the deformation of the microstructure due to applied shear. Consistent with the nature of the system, this linear viscoelastic measurement demonstrates the importance of the inter-particle interaction in controlling the rheological properties. The results of steady shear measurements, which were obtained simultaneously with the SANS experiments, are given in Fig. 1(b). Pronounced shear thinning behavior, along with a steady increase of stress, is observed within the accessed range of shear rate. It should be noted that the shear viscosity of this charged colloidal suspension exceeds that of the hosting solvent (0.021 Pa·s), by several orders of magnitude. Such solid-like rheological properties in oscillatory shear ($G' > G''$) and strong shear thinning behavior indicate the highly cooperative, glassy nature of our system.

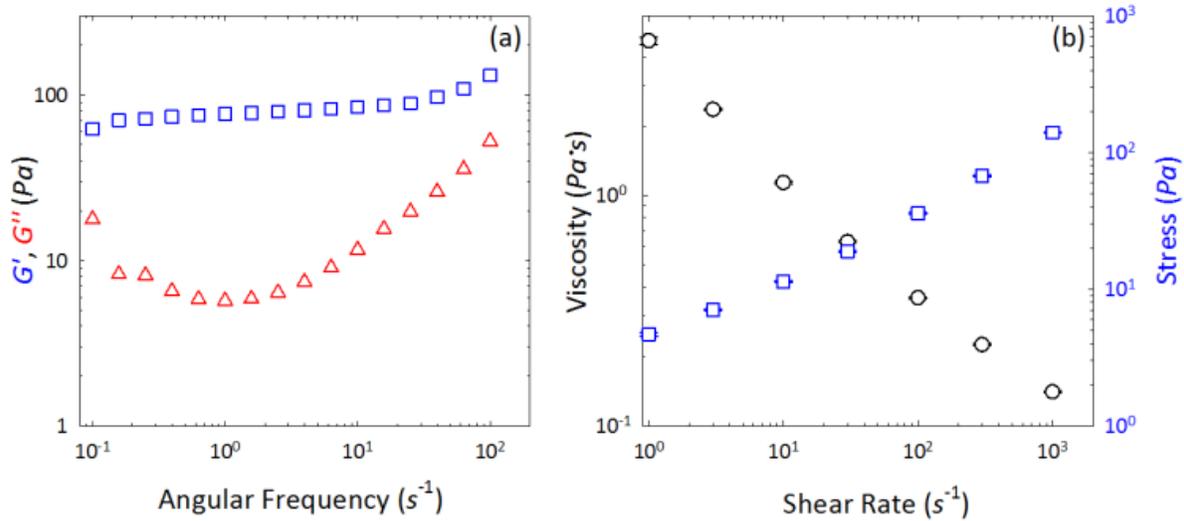

**Fig. 1** Rheological behavior of the charged colloidal suspension studied in this report. (a) Frequency dependence of the storage and loss moduli G' (blue symbols) and G" (red synbols). (b) Shear stress (blue symbols) and viscosity (black symbols) as a function of shear rate.

Small-angle neutron scattering (SANS) measurements were carried out at HFIR ORNL using the CG2 SANS spectrometer with a flow-vorticity (1-3) configuration.[33] The wavelength of the incident neutron beam was chosen to be 12 Å, with a wavelength spread $\frac{\Delta \lambda}{\lambda}$ of 15%, to cover values of the scattering wave vector $Q$ ranging from 0.001 to 0.02 Å$^{-1}$. The measured intensity $I(Q)$ was corrected for detector background and sensitivity, as well as the scattering contribution from the empty cell. The temperature was set to 20°C and a Couette shear cell was used to investigate the structure of flowing colloids projected onto the $1-3$ plane. The accessed range of shear rate was from 1 to 300 s$^{-1}$.

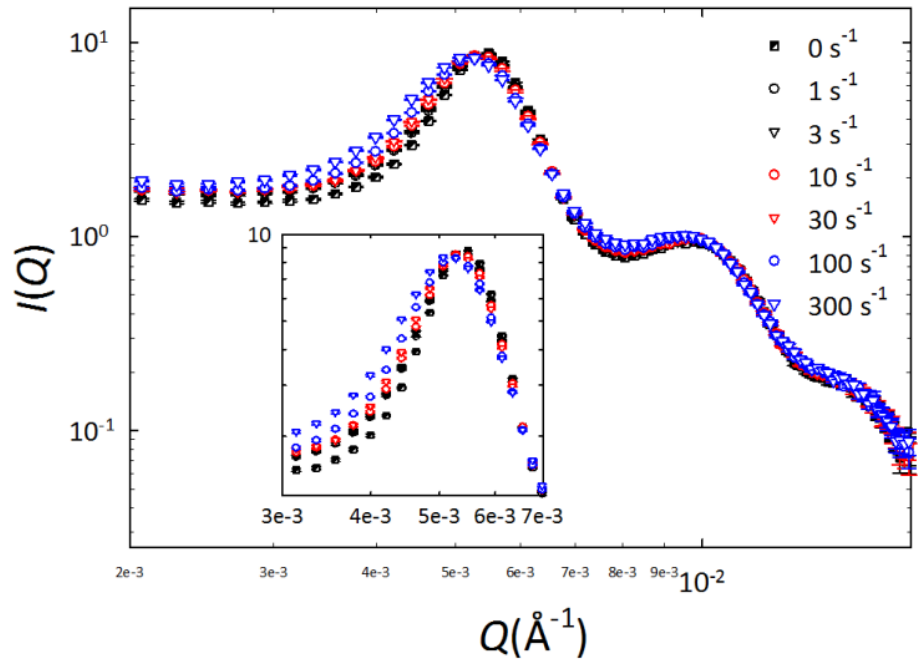

**Fig. 2** The comparison of the equilibrium $I(Q)$ ($\dot{\gamma} = 0$) and $I_0^0(Q)$, the isotopic component of the measured $I(\mathbf{Q})$, obtained at different shear rates. The results around the $Q$ range of the first interaction peak is enlarged and given in the inset.

When a macroscopic strain is applied to an amorphous material, the constituent particles are displaced from their equilibrium positions. It has been now recognized that such particle rearrangements in a highly cooperative colloidal system generally do not follow the macroscopic strain but are highly heterogeneous in space and time. Experimentally how to characterize this heterogeneous deformation in the nonequilibrium state remains a central problem. We should note that for materials subject to applied stress, the radial spatial distribution no longer provides a satisfactory description of the structure of constituent particles in the non-equilibrium state.[31] A complete description of inter-particle spatial arrangement requires both the radial and angular correlations. Recently we have formulated the equation[32] for extracting microscopically heterogeneous strain of deformed materials from scattering and computational studies by using a perturbation expansion of the anisotropic structure factor $S(\mathbf{Q})$ in terms of real spherical harmonics: First we expand the $S(\mathbf{Q})$ using the real spherical harmonic functions $Y_l^m\left(\frac{\mathbf{Q}}{|Q|}\right)$,

$$S(\mathbf{Q}) = \sum_{l,m} S_l^m(Q) Y_l^m\left(\frac{\mathbf{Q}}{|Q|}\right), \quad (1)$$

where $S_l^m(Q)$ are the expansion coefficients associated with the specific $Y_l^m\left(\frac{\mathbf{Q}}{|Q|}\right)$ with order $l$ and degree $m$. For materials subject to steady shear, we show that the connection between the spectral distortion and strain heterogeneity is given by:

$$S_0^0(Q) - S(Q) = \frac{2\pi\rho}{15} Q \int_0^\infty \gamma^2(r) g(r) \left[\frac{2r^2}{Q}\cos Qr - \left(\frac{2r}{Q^2} + r^3\right)\sin Qr\right] dr, \quad (2)$$

where $S(Q)$ and $g(r)$ are the inter-particle structure factor and pair distribution function of materials at their quiescent states, $S_0^0(Q)$ is the isotropic component of the anisotropic structure factor at different shear conditions extracted from the measured scattering signal on the 1-3 plane and $\gamma(r)$ is a function which provides the quantitative description about the local strain distribution. The resulting formula, which is independent of detailed inter-particle potential form, provides the mathematical basis for extracting microscopic deformation from scattering experiments and non-equilibrium molecular dynamics simulations of deformed materials. From analyzing the spectral distortion given by the difference between $S_0^0(Q)$ and $S(Q)$, $\gamma(r)$ can be determined accordingly.

Before we proceed further into the discussion of strain heterogeneity determined from our experiment, it is worth commenting on the dependence of $S_0^0(Q)$ on shear rate: Based on our previously developed method, the isotropic component of the scattering intensity $I_0^0(Q) \sim S_0^0(Q)$ can be extracted from the measured $I(\boldsymbol{Q})$ and the results are given in Fig. 2. In the quiescent state the coherent component of $I(Q)$ of a colloidal suspension consisted of spherical objects is proportional to the product of $S(Q)$ and the form factor $P(Q)$. Because of the rigidity of silica particles, $P(Q)$ of our system remains unchanged within the range of shear rate accessed in this study. Therefore $I_0^0(Q)$ extracted from $I(\boldsymbol{Q})$ at different shear rates retains the same qualitative features of $S_0^0(Q)$. Upon increase of the shear rate, the $I_0^0(Q)$ is characterized by a steady shift of its first interaction peak toward the lower-$Q$ region and a continuous increase in its zero-angle limit. These observations suggest the deteriorating local ordering and signify an increasing compressibility. These observations indicate the colloidal suspension is progressively fluidized by the imposed steady shear due to the increase in effective temperature predicted by earlier studies.[34]

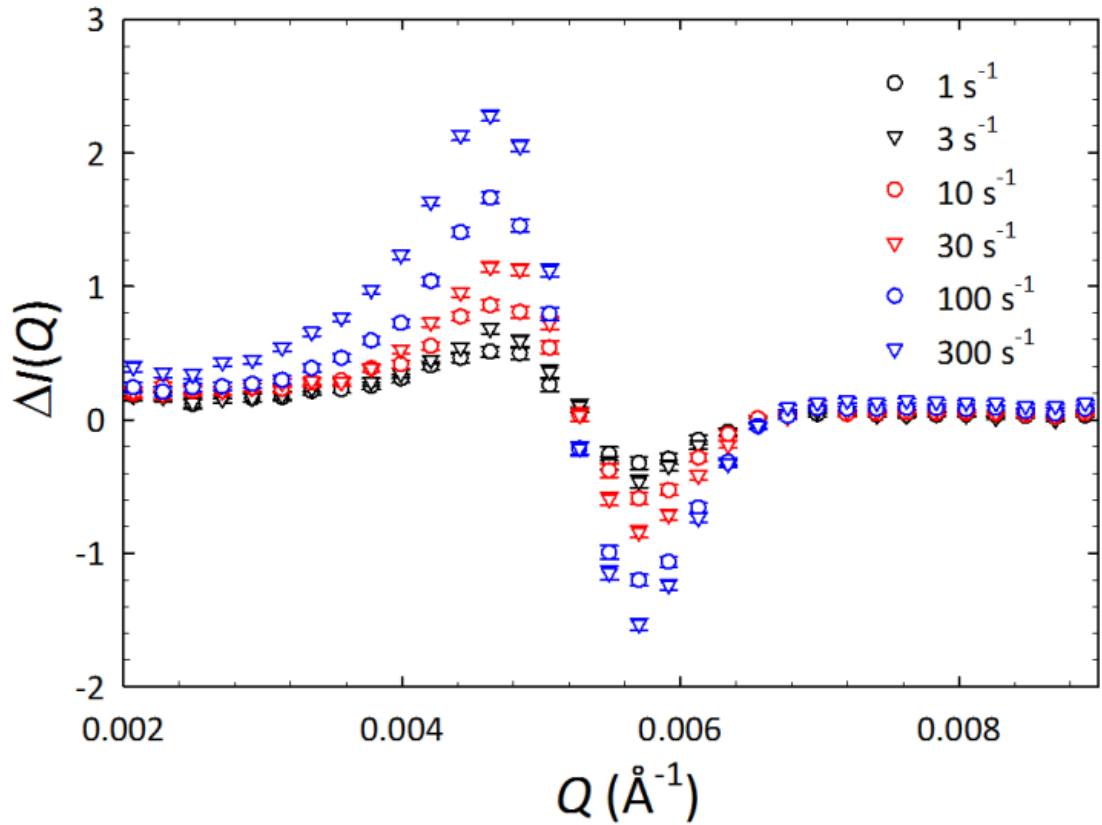

**Fig. 3** The $\Delta I(Q)$ obtained at different shear rate. The incresing magnitude of characteristic oscillations indicate the progressive configurational deviation of particles from their equilibrium positions.

Because $\gamma(r)$ is embedded in the integrand on the right hand side of Eqn. (2), its quantitative determination requires a data fitting of scattering results based on an explicit expression of $\gamma(r)$. While for our colloidal suspension a mathematical model of $\gamma(r)$ and its dependence on shear rate remain to be identified, Eqn. (2) nevertheless presents a framework which allows a qualitative, *model-free* examination of flow inhomogeneity from the anisotropic scattering spectra: If a system undergoes an affine deformation, the characteristic dependence of strain on $r$ no longer exists. In this ideal condition, the $\gamma(r)$ in Eqn. (2) can be simply replaced by a constant strain $\gamma$, which can be demonstrated to take the following expression:

$$\gamma = \sqrt{\frac{30[S_0^0(Q) - S(Q)]}{4Q\frac{dS(Q)}{dQ} + Q^2\frac{d^2S(Q)}{dQ^2}}}, \quad (3)$$

In Fig. 3 we present $\Delta I(Q) \equiv I_0^0(Q) - I(Q)$ as a function of shear rate. Upon increasing $\dot{\gamma}$, the characteristic oscillations of $\Delta I(Q)$, which reflects the same qualitative features of $\Delta S(Q) \equiv S_0^0(Q) - S(Q)$ as we just point out earlier, is seen to become increasingly pronounced. Based on its definition, it is clear that $\Delta S(Q)$ gives the statistical average of the shear-induced microstructural distortion from the equilibrium positions. From the dependence of $\Delta I(Q)$ on shear rate, the micromechanical behavior of flowing colloids can be pictorially envisioned from a perspective of local topology: Existing studies have clearly demonstrated that amorphous materials are generally characterized by highly heterogeneous local elastic modulus originating from the widely different local configurational environment.[13,35-39] When the applied shear stress is small, the majority of particles is slightly driven out of their equilibrium configuration by a relatively small strain. During this process the loaded local stress is insufficient to overcome the energy barrier to cause significant topological variations. Namely the microstructural distortion is characterized by a nearly affine deformation. On the contrary, when the system is subject to a large macroscopic shear stress with imposing mechanical energy which exceeds the threshold set by the local modulus, more particles therefore can now be freed from their immediate surroundings via plastic rearrangements. This deformation behavior is accompanied by a significant deviation from the equilibrium configuration, characterized by larger strains and reflected by a more pronounced variation in $\Delta I(Q)$. It is therefore clear that, depending on the interplay of thermodynamical and fluid mechanical interactions in the steady states at different shear rates, the anisotropic scattering intensity contains the structural information from both affine and plastic deformations in varying amounts. However judging from $\Delta I(Q)$ at a given shear rate, whether the colloidal motion is dominated by affine or plastic dynamics cannot be directly identified.

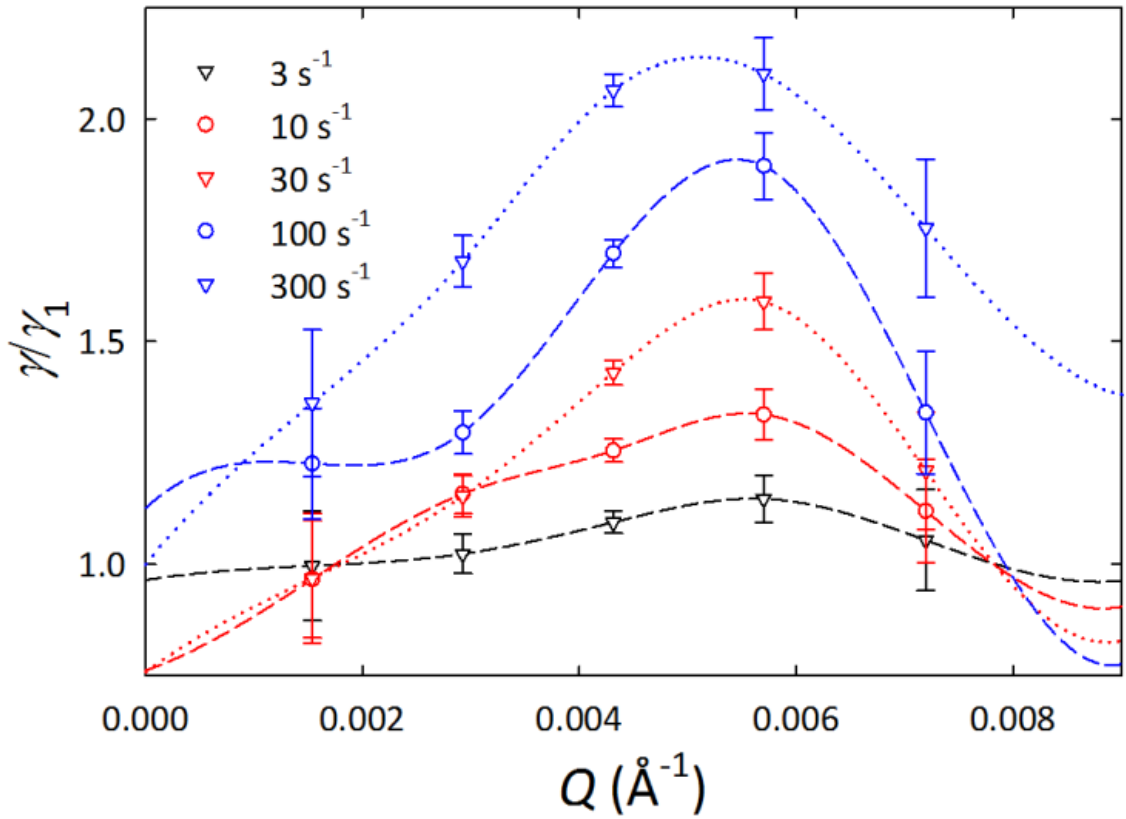

**Fig. 4** The ratio of $\gamma$ obtained experimentally based on Eqn. (3) and that of $\dot{\gamma} = 1$. Symbols are the experimental results and the lines are used to guide the reader's eye.

To resolve this ambiguity we present the square root of the ratio between $\Delta I(Q)$ at a given shear rate to that corresponding to $\dot{\gamma} = 1\ \text{s}^{-1}$ in Fig. 4. From Eqn. (3), it can be seen that this quantity is indeed equal to $\frac{\gamma}{\gamma_1}$ because the factor of $\frac{30}{4Q\frac{dS(Q)}{dQ}+Q^2\frac{d^2 S(Q)}{dQ^2}}$ and $P(Q)$ are eliminated in this operation. In the case of $\dot{\gamma} = 1\ s^{-1}$, the particle configuration is only slightly driven out of the equilibrium conditions due to the small mechanical perturbation. As a result, the particle deformation should be characterized by a nearly affine behavior and we therefore select $\gamma_1$ as a yardstick, i.e., a normalization factor, to evaluate the deformation behavior of different shear rates. Upon increasing $\dot{\gamma}$, a developing broad peak in $\frac{\gamma}{\gamma_1}$ centered between $Q = 0.004\ \text{Å}^{-1}$ and $0.006\ \text{Å}^{-1}$ is clearly revealed. This observed progressive deviation from the $Q$-independent constant value of $\frac{\gamma}{\gamma_1}$ clearly evidences a transition in the deformation mechanism from affine behavior to plastic flows from analysis of the spectral anisotropy.

## Conclusions

To summarize, we have shown that our developed method provides an adequate and convenient experimental tool for qualitatively evaluating the strain heterogeneity in flowing colloids from the experimental anisotropic scattering spectra. It should facilitate the investigation of the non-equilibrium structure of general soft matter materials. Our method allows for a more direct characterization of strain heterogeneity predicted by theory and simulations, and its application could impact and benefit the research in the larger field of material deformation in soft matter, e.g., biological topics such as the flow mechanism and viscoelasticity of blood, and deformation characteristics in hard matter like asphalt, cements, and metal alloys which are relevant to many industrial applications. It is in principle also possible to extend the current method to quantitatively examine flow heterogeneity at the particle level. An implementation of this approach based on Eqn. (2), which is beyond the scope of this report, is currently under development.

## Acknowledgements

This research at CNMS and SNS of Oak Ridge National Laboratory was sponsored by the Scientific User Facilities Division, Office of Basic Energy Sciences, U.S. Department of Energy. K.C. acknowledges the support of ORNL-HSRE program. G.Z. thanks the support of ORNL-HERE program. G.-R.H. acknowledges the supports from the National Center for Theoretical Sciences, Ministry of Science and Technology in Taiwan (Project No. MOST 106-2119-M-007-019) and the Shull Wollan Center during his visit to ORNL.

## Conflicts of interest

There are no conflicts to declare.

## Notes and references

1 *Colloidal Dispersions*, W. B. Russel, D. A. Saville and W. R. Schowalter, Cambridge University Press, Cambridge, 1989.

2 *The Structure and Rheology of Complex Fluids*, R. G. Larson, Oxford University Press, New York, 1999.

3  *Colloidal Suspension Rheology,* J. Mewis and N. J. Wagner, Cambridge University Press, Cambridge, 2012.

4  J. F. Morris, *Rheol. Acta,* 2009, **48,** 909.

5  D. R. Foss and J. F. Brady, *J. Fluid Mech.,* 2000, **407,** 167.

6  J. R. Melrose and R. C. Ball, *J. Rheol.,* 2004, **48,** 937.

7  J. Bergenholtz, *Curr. Opin. Colloid Interface Sci.,* 2001, **6,** 484.

8  J. M. Brader, *J. Phys.: Condens. Matter,* 2010, **22,** 363101.

9  T. Voigtmann, *Curr. Opin. Colloid Interface Sci.,* 2014, **19,** 549.

10 R. A. Lionberger, *J. Rheol.,* 1998, **42,** 843.

11 J. M. Brader, M. E. Cates and M. Fuchs, *Phys. Rev. Lett.,* 2008, **101,** 138301.

12 E. Nazockdast and J. F. Morris, *J. Fluid Mech.,* 2012, **713,** 420.

13 Dynamical Heterogeneities in Glasses, Colloids, and Granular Media, ed. L. Berthier, G. Biroli, J.-P. Bouchaud, L. Cipelletti and W. van Saarloos, Oxford University Press, Oxford, 2011.

14 R. Yamamoto and A. Onuki, *Europhys. Lett.,* 1997, **40,** 61.

15 R. Yamamoto and A. Onuki, *Phys. Rev. E,* 1998, **58,** 3115.

16 K. Miyazaki, D. R. Reichman and R. Yamamoto, *Phys. Rev. E,* 2004, **70,** 011501.

17 A. Furukawa, K. Kim, S. Saito and H. Tanaka, *Phys. Rev. Lett.,* 2009, **102,** 106001.

18 D. Chen, D. Semwogerere, J. Sato, V. Breedveld and E. R. Weeks, *Phys. Rev. E,* 2010, **81,** 011403.

19 J. Clara-Rahola, T. A. Brzinski, D. Semwogerere, K. Feitosa, J. C. Crocker, J. Sato, V. Breedveld and E. R. Weeks, *Phys. Rev. E,* 2015, **91,** 010301(R).

20 R. Besseling, E. R. Weeks, A. B. Schofeld and W. C. K. Poon, *Phys. Rev. Lett.,* 2007, **99,** 028301.

21 V. Chikkadi, G. Wegdam, D. Bonn, B. Nienhuis and P. Schall, *Phys. Rev. Lett.,* 2011, **107,** 198303.

22 V. Chikkadi, S. Mandal, B. Nienhuis, D. Raabe, F. Varnik and P. Schall, *Europhys. Letts.,* 2012, **100,** 56001.

23 V. Chikkadi and P. Schall, *Phys. Rev. E,* 2012, **85,** 031402.

24 S. Mandal, V. Chikkadi, B. Nienhuis, D. Raabe, P. Schall and F. Varnik, *Phys. Rev. E,* 2013, **88,** 022129.

25 F. Varnik, S. Mandal, V. Chikkadi, D. Denisov, P. Olsson, D. VAppl. Opt.gberg, D. Raabe and P. Schall, *Phys. Rev. E,* 2014, **89,** 040301(R).

26 V. Chikkadi, D. M. Miedema, M. T. Dang, B. Nienhuis and P. Schall, *Phys. Rev. Lett.,* 2014, **113,** 208301.

27 N. A. Clark and B. J. Ackerson, *Phys. Rev. Lett.,* 1980, **44,** 1005.

28 S. J. Johnson, C. G. de Kruif and R. P. May, *J. Chem. Phys.,* 1988, **89,** 5909.

29 C. P. Amann, D. Denisov, M. T. Dang, B. Struth, P. Schall and M. Fuchs, *J. Chem. Phys.,* 2015, **143,** 034505.

30 A. K. Gurnon and N. J. Wagner, *J. Fluid Mech.,* 2015, **769,** 242.

31 J. Vermant, and M. J. Soloman, *J. Phys.: Condens. Matter,* 2005, **17,** R187.

32 G.-R. Huang, B. Wu, Y. Wang and W.-R. Chen, submitted. Manuscript is available from arXiv:1709.07507 [cond-mat.soft].

33 L. Porcar, D. Pozzo, G. Langenbucher, J. Moyer and P. D. Butler, *Rev. Sci. Instrum.,* 2011, **82,** 083902.

34 M. L. Falk and J. S. Langer, *Phys. Rev. E.,* 1998, **57,** 7192.

35 C. Maloney and A. Lemâtre, *Phys. Rev. Lett.,* 2004, 93, 016001.

36 C. Maloney and A. Lemâtre, *Phys. Rev. Lett.,* 2004, 93, 195501.

37 N. P. Bailey, J. Schiøtz, A. Lemâtre and K. W. Jacobsen, *Phys. Rev. Lett.,* 2007, **98,** 095501.

38 A. Lemâtre and C. Caroli, *Phys. Rev. Lett.,* 2009, **103,** 065501.

39 H. Mizuno, S. Mossa and J.-L. Barrat, *Europhys. Lett.,* 2013, **104,** 56001.